\documentstyle[12pt]{article}

\newcommand{\nc}{\newcommand}
\nc{\beq}{\begin{equation}}
\nc{\eeq}{\end{equation}}
\nc{\beqa}{\begin{eqnarray}}
\nc{\eeqa}{\end{eqnarray}}

\input epsf
\newwrite\ffile\global\newcount\figno \global\figno=1

\def\writedef#1{}
\def\figin{\epsfcheck\figin}\def\figins{\epsfcheck\figins}
\def\epsfcheck{\ifx\epsfbox\UnDeFiNeD
\message{(NO epsf.tex, FIGURES WILL BE IGNORED)}
\gdef\figin##1{\vskip2in}\gdef\figins##1{\hskip.5in}
\else\message{(FIGURES WILL BE INCLUDED)}%
\gdef\figin##1{##1}\gdef\figins##1{##1}\fi}
\def\figinsert{}
\def\ifig#1#2#3{\xdef#1{fig.~\the\figno}
\writedef{#1\leftbracket fig.\noexpand~\the\figno}%
\figinsert\figin{\centerline{#3}}\medskip\centerline{\vbox{\baselineskip12pt
\advance\hsize by -1truein\center\footnotesize{  Fig.~\the\figno.} #2}}
\bigskip\endinsert\global\advance\figno by1}
\def\endinsert{}

\begin{document}

\title{\large{\bf Zero Energy Configurations in General Relativity}}

\author{
Stephen D.H.~Hsu\thanks{Address after January, 1998: Dept. of Physics,
University of Oregon, Eugene OR 94703-5203. Email:
hsu@duende.uoregon.edu}
\\ \\ Department of Physics, Yale University, New Haven, CT 06520-8120 \\
}

\date{January, 1997}

\maketitle

\begin{picture}(0,0)(0,0)
\put(350,290){YCTP-P29-97}
\end{picture}
\vspace{-24pt}

\begin{abstract}
We investigate the ratio of gravitational binding energy 
to rest mass in general relativity. For N pointlike masses,
an upper bound on the magnitude of this ratio can be derived 
using the second
law of black hole dynamics. Only as N approaches infinity
can it approach one. A configuration that saturates the
${\rm N} = \infty$ bound is a thin spherical shell as its radius is
taken to zero. This system provides, in principle, a perfectly
efficient mechanism for converting rest mass into energy. We mention
possible implications for the black hole information problem.

\end{abstract}

\newpage


Consider two pointlike objects of mass $m$ separated 
by a distance $r$. In Newtonian
gravity, the total energy of the system can 
be made as large and negative
as we like by taking $r$ to zero. In general 
relativity (GR), positivity
theorems \cite{positive} preclude this possibility, 
but it remains interesting
to ask whether the total energy of systems of this 
type can ever reach zero.
In other words, can the gravitational binding 
energy ever be sufficient to
cancel the rest masses of the particles? 
This is analogous to a question which
fascinates particle physicists: when are microscopic 
(e.g. gauge) forces sufficient to create
massless bound states out of massive constituents? 
In the case of two
point masses in GR there is a lower limit on $r$ 
due to the formation of a horizon
which is of order $r_s \sim m$ (here we 
adopt units in which Newton's constant 
$G_N = 1$). Substituting $r_s$ into the 
Newtonian formula for the binding energy
one obtains
$$ U(r_s) ~=~ - m^2 / r_s ~\sim~  - m ~~,$$
so we expect that the maximum binding energy 
is of the same order of 
magnitude as $m$.

We recall that while the definition
of energy in GR can be subtle \cite{Wald}, 
in the case of an asymptotically
flat universe there is a natural definition 
which coincides with the gravitational
effect of the bound system on a test mass 
placed far away in the asymptotic region.
Thus a zero energy bound system has no 
gravitational effect on a test charge at infinity
and causes no spacetime curvature at large distances.

It is straightforward to derive a lower bound on the 
energy of a system constructed from
N pointlike objects of mass $m$. In GR the 
closest one can come
to a pointlike mass is a black hole. 
Bringing the black holes sufficiently close
 together we can merge them together to form
a single, larger black hole of mass $M$. 
(For simplicity, we assume zero
total charge and angular momentum.)
Due to Hawking's second law of 
black hole dynamics
\cite{Hawking}, which states that the total 
area of the black holes must never decrease, 
we know that
$$ M^2 \geq N m^2 ~~.$$
This implies that $M \geq \sqrt{N} m$, and 
yields an upper bound on the magnitude of the
binding energy:
$$ |U| ~\leq~ N m ~( 1 - 1 / \sqrt{N} ) ~~.$$
We see that a zero energy bound system can 
only be achieved in the ${\rm N} \rightarrow \infty$
limit.

There is a simple example of a system which saturates 
the zero energy bound. Consider a
thin spherical shell of radius $r$ and mass $M_0$. 
(We can imagine that it is constructed out of N 
objects uniformly distributed in a spherically 
symmetric way at radius $r$.) 
We can compute its mass using the standard 
results for the solution of Einstein's equation
with spherical symmetry \cite{Wald}. We find 
that the total energy of the system obeys the
relation 
$$ E ~=~ M_0 \sqrt{ 1 - 2 E / r } ~~.$$
The factor $\sqrt{ 1 - 2 E / r }$ arises because the proper three-volume 
for the interior Schwarzschild solution is given by
$$
\sqrt{ {}^3 g} ~d^3x ~=~ {r^2 \over \sqrt{ 1 - 2 E / r } }~dr~ d \Omega ~~.
$$
Solving for $E / M_0$, we obtain
$$ E / M_0 ~=~ ( 1 + M_0^2 / r^2 )^{1/2} ~-~ M_0 / r ~~.$$
For large $r$ this ratio is close to unity, while as 
$r \rightarrow 0$ it approaches zero.
Note that this system is not obtained 
by the free collapse of a spherical shell. In that case
the potential energy is simply converted to 
kinetic energy and the total energy (hence
the gravitational effect at infinity) remains 
constant at $E = M_0$. In order to construct
the system described above, we need to shrink 
the shell quasi-statically, extracting all of
the kinetic energy liberated by collapse. 
A mechanism for doing so, such as resilient cables attached to
each of the N elements of the shell, would be difficult to implement.
Indeed, one can see that as $r \rightarrow 0$ the total
tension on the N cables would be of Planckian size, perhaps
necessitating actual cosmic strings as cables.
Also, we have neglected all quantum
gravitational effects, such as a minimum allowed
thickness of the shell which would be of order the
Planck length. For a shell of finite thickness the
limiting value of $E$ would be non-zero.

It is worth noting that \cite{Lightman} 
for a point particle the energy, given by the dot product of
its four-momentum and the time Killing vector
$$ E ~=~ {\bf p \cdot \xi } ~~,$$ 
obeys the bound
$$ E ~\geq~ m  ~( {\bf \xi \cdot \xi})^{1/2} ~~$$
when $\xi$ is timelike. ($\bf \xi$ can become spacelike, 
for example inside the ergosphere of
a black hole). Thus, $E$ is positive except where 
the norm of $\bf \xi$ changes sign
-- i.e. at an event horizon. It is easy to see 
that in the spherical shell system 
the particles in the shell are approaching 
a zero-radius horizon as 
$r$ goes to zero.

It is amusing that the spherical 
shell system allows for the
possibility of extremely efficient conversion of rest 
mass into energy. A ``sufficently
advanced'' civilization could construct such an 
object (an abandoned Dyson sphere?) and
allow it to collapse quasi-statically with extremely 
light and resilient strings attached 
to each of the N components. If the strings are 
connected to engines which maintain the 
components at finite velocity during the collapse, 
the liberated kinetic energy could be extracted and utilized.
At the end of the collapse one is left with a 
very small black hole ($E << M_0$) which 
then evaporates. Note that the requirements on the
properties of the strings, such as the maximum
tension they can sustain, decrease with increasing N.

Finally, we mention the possible implications of
the spherical shell system for the black hole information
problem. The construction described here allows a very large
initial object, potentially containing a vast amount of
information, to be hidden in an extremely light, almost
zero-radius black hole. If information could be stored this
way, the usual interpretation of area as entropy would be
problematic.
What is not clear is to what extent correlations
between the black hole and the state of the external
universe are induced during the quasi-static collapse.
If they are, the information is not truly hidden in the black hole, 
but can be reconstructed from the exterior state. In the usual
thought experiment, in which an object falls into an already
existing black hole, it is assumed that this is not the case
because the horizon of a sufficiently large black hole 
is an innocuous place and nothing special seems to be
occurring there. Here the answer is less clear because of the
required interactions between the shell and external forces which
slow the collapse.

\bigskip 
\noindent {\bf Note added:} After this work was completed, we
were made aware of several related investigations. That the thin
shell configuration has zero energy in the zero radius limit
was first discussed in \cite{ADM} (see also \cite{Malec}). 
Other zero energy configurations are discussed in
\cite{LH} and \cite{ZN}.

\bigskip
\noindent 
The author would like to thank James Hormuzdiar and Stephen Selipsky
for useful discussions and comments.
This work was supported in part under DOE contract DE-AC02-ERU3075.

\end{document}